\documentclass{article}
\usepackage{amsfonts}
\usepackage{amsmath}
\usepackage{epsfig}

\begin{document}

\title{\hspace{4.1in}{\small CERN-PH-TH/2004-259; OUTP-0502P}\\
\vspace*{1cm} Racetrack inflation and assisted moduli stabilisation}
\author{\bigskip Z. Lalak$^{a, b}$, G. G. Ross$^{a, c}$ and S. Sarkar$^{c}$\\
$^{a}$ Theory Division, CERN, 1211 Geneva 23, Switzerland\\
$^{b}$ Institute of Theoretical Physics, University of Warsaw, 00-681
       Warsaw, Poland\\
$^{c}$ Rudolf Peierls Centre for Theoretical Physics,\\
       University of Oxford, 1 Keble Road, Oxford, OX1 3NP}
\date{}
\maketitle

\begin{abstract}
We present a model of inflation based on a racetrack model
\emph{without} flux stabilization. The initial conditions are set
automatically through topological inflation. This ensures that the
dilaton is not swept to weak coupling through either thermal effects
or fast roll. Including the effect of non-dilaton fields we find that
moduli provide natural candidates for the inflaton. The resulting
potential generates slow-roll inflation without the need to fine-tune
parameters. The energy scale of inflation must be near the GUT scale
and the scalar density perturbation generated has a spectrum
consistent with WMAP data.
\end{abstract}

\bigskip

\section{Introduction}

The issues of hierarchical supersymmetry breakdown and moduli
stabilization have long been the central problems of superstring
phenomenology. In the original version of the weakly coupled heterotic
string these two phenomena are related to each other --- once
supersymmetry is broken a potential for the moduli fields is generated
and their expectation values, hence the parameters of the low energy
effective Lagrangian, are determined. If one is to quantify this, the
first question to answer is what is the source of supersymmetry
breakdown? The next task is to explain how some of the (local or
absolute) vacuua of the moduli have just those values which correspond
to the observed Universe. Finally one needs to explain how, given that
there are many candidate vacuua, just the right one is selected. This
latter problem turns out to be particularly difficult to solve because
in string-motivated models there always exists a trivial vacuum
corresponding to a noninteracting theory. The problem arises because
this trivial vacuum is separated from the relevant non-trivial vacuua
by an energy barrier which is many orders of magnitude lower than the
Planck scale; this follows because of the large hierarchy between the
Planck scale and the supersymmetry breaking scale. The problem is that
if the initial values for moduli fields are not fine-tuned, the moduli
roll so fast that they readily cross the barrier separating the
desired minimum from the trivial vacuum. Thus the problem of moduli
stabilization is not simply a question of the existence of suitable
minima of the potential, it is rather a question about the
cosmological dynamics of the theory. These problems have been
discussed many times over the last two decades without a satisfactory
resolution, see e.g. ref.\cite{Banks:1995dp}.

Recently a new twist has been added to the discussion in the context
of type IIB string compactifications with fluxes. It has been shown
that due to the presence of nonzero fluxes of form-fields, all but one
of the moduli, including the dilaton but excluding the overall
breathing mode of the compact manifold, can be stabilised
\cite{Kachru:2003aw,Kallosh:2004yh}. The remaining volume modulus can
be stabilised with the help of brane-antibrane forces combined with
the gaugino-condensation induced nonperturbative
potential.~\footnote{However it may be necessary to re-examine the
KKLT stabilization of the dilaton in the presence of the
superpotential fot the $T$ modulus \cite{Choi:2004sx}.} Moreover, in
this context the axionic superpartner of the volume modulus can play
the role of the cosmological inflaton. To achieve this the authors of
ref.\cite{Blanco-Pillado:2004ns} invoked a constant piece in the
superpotential and the explicitly supersymmetry-breaking term in the
scalar potential associated with brane-antibrane forces. However to
achieve a significant amount of inflation it proved necessary to
fine-tune parameters at the level of one part in a thousand. It was
argued on anthropic grounds that such fine-tuning is acceptable. Here
we eschew such arguments and look for inflationary solutions which do
not involve any fine-tuning --- for other relevant proposals see
refs.\cite{Brustein:2002mp,Iizuka:2004ct}.

In this paper we consider these problems in the more traditional setup
of `racetrack' models having two gaugino condensates, generated by two
asymptotically free non-Abelian hidden gauge sectors in $N=1$
supergravity
\cite{Dixon:1990ds,Krasnikov:1987jj,Casas:1990qi,deCarlos:1992da}. Unlike
the fluxed models we do not include a constant piece in the
superpotential, or break supersymmetry explicitly. We give special
attention to the trapping, through topological inflation, of the
dilaton at a finite value away from the zero coupling limit at
infinity. We discuss how this can avoid the fast-roll problem, and
also the problem associated with thermal corrections to the dilaton
potential. Although we concentrate on the implications for the weakly
coupled heterotic string without fluxes, our results are also relevant
to type IIB string compactifications with fluxes. We stress that all
this is possible within the well known field-theoretical framework of
gaugino-condensation racetrack models; hence the mechanism should be
applicable in a wide range of models obtained from various string
compactifications.

In Section~2 we discuss how the racetrack potential, through threshold
effects, necessarily involves non-dilaton fields, in particular
moduli. We discuss how these effects are included in the
superpotential and calculate the resulting scalar potential.

In Section~3 we consider how inflation can occur through the slow-roll
relaxation of such fields. We show that while slow-roll in the dilaton
direction can occur after inflation, the theory is likely to be driven
to the free limit. This problem does not occur for the inflationary
periods generated by slow-roll in the moduli direction. In this case
there is a pseudo-Goldstone inflaton which naturally solves the
`$\eta$ problem'. We present a numerical study of slow-roll inflation
in which we show that the model readily leads to an adequate period of
inflation with density perturbations in agreement with observation and
a spectral index between 0.96 and 1 (the allowed $1\sigma$ range from
WMAP-3 being $0.932 \leq n_{s} \leq 0.970$, if a scale-free power-law
spectrum is assumed \cite{Spergel:2006hy}).

In Section~4 we discuss the initial conditions for inflation. We show
how the race-track potential necessarily exhibits topological
inflation which automatically sets the appropriate initial conditions
leading to the slow-roll inflationary era discussed in Section~3. In
particular, provided the reheat temperature after inflation is not too
large, topological inflation provides a simple mechanism to evade the
thermal overshoot problem \cite{Buchmuller:2004xr} and avoid the
dilaton overshoot problem \cite{Brustein:1992nk}. We also discuss how
the non-inflaton moduli fields are stabilised in a manner that avoids
the difficulties discussed in ref.~\cite{Choi:2004sx}. Finally in
Section~5 we present a summary and our conclusions.

\section{The gaugino condensation superpotential}

There are notable differences between gaugino condensation
\cite{Taylor:1982bp,Dine:1985rz,Derendinger:1985kk,Amati:1988ft} and
fluxes as sources of the superpotential. Perhaps the most important
for the present discussion is the fact that the gaugino condensation
scale is sensitive to the beta function which in turn is sensitive to
the actual mass spectrum at a given energy scale, so the effective
nonperturbative potential responds to expectation values of various
fields which control the actual mass terms.  Fluxes, on the other
hand, are imprinted into geometry of the compact manifold and the
parameters of the respective superpotential are fixed below the Planck
scale. In this paper we will show that the fact that the gaugino
condensates take into account the backreaction of other fields on the
dilaton provides a natural origin for an inflaton sector. We begin
with a discussion of this important backreaction effect.

\subsection{The condensation scale}

The relation between the field theoretical coupling and the string
coupling at the Planck scale is given by
\begin{equation}
\frac{4 \pi}{g_i^2 (M)} = \frac{4\pi}{g_{\mathrm{s}}^2} 
 + \frac{\Delta_i}{4 \pi}\ ,  
\label{def1}
\end{equation}
where $\Delta_i$ denote string threshold corrections computed at the
string scale. The string coupling is related to the dilaton field by
$\mathrm{Re} (S) = 2\pi/\alpha_{\mathrm{s}}$. At 1-loop the running
coupling is given by
\begin{equation}
\frac{1}{\alpha (Q)} = \frac{1}{\alpha (\mu)} + 
 \frac{b_0^\prime}{2 \pi} \log \left(\frac{Q}{\mu}\right) ,  
\label{running}
\end{equation}
where for supersymmetric $SU (N)$ with $K$ copies of the vectorlike
representation $N + \bar{N}$, the beta function coefficient
$b_0^\prime$ is given by: $b_0^\prime = 3N - K$. The gaugino
condensation scale, $\Lambda$, is given approximately by the scale $Q$
at which $1/\alpha (Q) \rightarrow 0$. Using this definition one
obtains
\begin{equation}
\Lambda = M \mathrm{e}^{-\mathrm{Re} (S)/b_0^\prime} ,  
\label{condensate}
\end{equation}
where $M$ is the string cut-off scale, which in leading order
corresponds to the gauge coupling unification scale ($5 \times
10^{17}$ GeV in the weakly-coupled heterotic string).

The effect of string threshold corrections can be estimated in field
theory by including the contribution to the condensation scale from
heavy states with mass less than the string cut-off scale, which is
typically close to the Planck scale. When the compactification radius
is greater than the string radius, the massive fields include those
Kaluza-Klein modes lighter than the string scale, but there may be
further massive fields which get their mass through a stage of
symmetry breaking after compactification.

Now if some gauge non-singlet fields acquire mass $m$ below the string
cut-off scale, then the beta function, $b_0^\prime$, changes to
$b_0$. In this case the condensation scale acquires an explicit
dependence on the scale $m$:
\begin{equation}
\Lambda = M \mathrm{e}^{-\mathrm{Re}(S)/b_0} 
 \left\vert\frac{M}{m}\right\vert^{(b_0^\prime - b_0)/b_0} .  
\label{fieldthreshold}
\end{equation}
We see that the 1-loop threshold correction is of the form
$\left\vert\frac{M}{m}\right\vert^{(b_0^\prime - b_0)/b_0}$ which
needs to be summed over the contribution of all massive states below
the cut-off scale.

\subsection{Moduli dependence of the gaugino condensate 
\label{modulidependence}}

In the case the mass, $m$, has a dependence on the vacuum expectation
value, $\langle\chi\rangle$, of a scalar field, $\chi$ it is clear
that the condensation scale, and hence the dilaton potential which
determines the condensation scale, will also depend on
$\langle\chi\rangle$. A particularly interesting case is the
dependence of the masses on moduli fields. This can arise through the
moduli dependence of the couplings involved in the mass generation
after a stage of symmetry breakdown. For example a gauge-non singlet
field, $\Phi$, may get a mass from a coupling to a field, $\Psi$, when
it acquires a vacuum expectation value through a Yukawa coupling in
the superpotential of the form $W = \lambda\Psi\Phi\Phi$.~\footnote{We
use the same symbol for the superfield and its $A$ component.} In
general the coupling $\lambda$ will depend on the complex structure
moduli, $\lambda = \lambda (\langle\chi\rangle)$, and so the mass of
the $\Phi$ field will be moduli dependent: $m = \lambda
(\langle\chi\rangle) \langle\Psi\rangle$. In what follows we will
consider the implications of a very simple dependence of $m$ on the
$\langle\chi\rangle$ which is sufficient to illustrate how $\chi$ can
provide an inflaton if it is a (modulus) field which has no potential
other than that coming from the $m$ dependence of the condensation
scale (and the above superpotential coupling). In particular we take
$m = \alpha + \beta\langle\chi\rangle$ where $\alpha$ is a mass coming
from another sector of the theory, possibly also through a stage of
symmetry breaking. With this we have $|m| = |\alpha + \beta\langle\chi
\rangle|$ in eq.(\ref{fieldthreshold}). We shall assume for simplicity
a canonical kinetic term for the modulus, $K=\bar{\chi}\chi$, although
in practice the form of its K\"{a}hler potential may be more
complicated.

\subsection{The moduli dependent superpotential}

Gaugino condensation is described by a nonperturbative superpotential
involving the dilaton superfield of the form $W_{\mathrm{npert}} = A
N_1 M^3 \mathrm{e}^{-S/N_1}$
\cite{Dine:1985rz,Derendinger:1985kk,Amati:1988ft}. The question that
arises at this point is how to turn the non-holomorphic threshold
dependence that enters the condensation scale (\ref{fieldthreshold})
due to the running in eq.(\ref{running}) into a holomorphic
superpotential. As we have discussed the radiative corrections giving
rise to the moduli dependence of the gauge coupling naturally involve
the modulus of the mass of the states involved in the loop. However
this is because one is calculating the Wilsonian coupling,
$g_{\mathrm{c}}$, with canonical kinetic terms, which is not
holomorphic. There is a simple relation of this coupling to the
holomorphic coupling, $g_{\mathrm{h}}$, with non-canonical kinetic
term \cite{Arkani-Hamed:1997mj} which is the quantity most closely
connected to the superpotential description of gaugino
condensation. At the 1-loop level one has
\begin{equation}
\mathrm{Re} \left(\frac{1}{g_{\mathrm{h}}^2}\right) =
 \frac{1}{g_{\mathrm{c}}^2}\ .  
\label{relation}
\end{equation}
Since the non-holomorphicity arises from the term proportional to
$\log \frac{M}{|\alpha + \beta \langle\chi\rangle|}$ on the rhs of
eq.(\ref{relation}), the only way of defining the holomorphic gauge
coupling which is consistent with eq.(\ref{relation}) is to replace
the absolute value $|\alpha + \beta\langle\chi\rangle|$ appearing
under the logarithm by $\alpha + \beta\chi$ itself (i.e. to restore
the dependence on the phase of $\chi$).  This leads to the holomorphic
gauge kinetic function below the scale set by $|\chi|$ of the form
\begin{equation}
f (S, \chi) = \frac{S}{8 \pi^2} - \frac{b_0^\prime - b_0}{8 \pi^2}
 \log\left(\frac{M}{\alpha + \beta \chi}\right) ,
\end{equation}
where the kinetic term of the gauge field has normalisation given by
$L_{\mathrm{k}} = \frac{1}{4} \mathrm{Re} (f(S, \chi))F^2$. As a
consequence, corresponding to the gaugino condensation scale
$\Lambda^3$ of eq.(\ref{condensate}) we have the holomorphic
superpotential
\begin{equation}
W = C M^3 \mathrm{e}^{-24 \pi^2 f (S, \chi)/b_0} ,  
\label{hcondensate}
\end{equation}
where the normalisation factor, $C$, is given in ref.\cite{Amati:1988ft}.

That this is the correct field dependence of the nonperturbative
superpotential can also be seen by considering the supergravity action
in its component form with explicit vacuum expectation values for the
gaugino bilinears. The relevant part of the scalar potential is
\begin{equation}
V = \mathrm{e}^{K} g^{i \bar{j}} \left(D_i W_{\mathrm{pert}} +
 \frac{1}{4} \mathrm{e}^{-K/2}\partial_{i} f
 \langle\lambda \lambda\rangle\right)
\left(D_{\bar{j}} \bar{W}_{\mathrm{pert}} + \frac{1}{4}
\mathrm{e}^{-K/2}\partial_{\bar{j}}\bar{f}
\langle\bar{\lambda} \bar{\lambda}\rangle\right) .
\end{equation}
Following Kaplunovski and Louis \cite{Kaplunovsky:1994fg}, we take 
\begin{equation*}
\langle\lambda \lambda \rangle_{\mathrm{sugra}} = 
\mathrm{e}^{K/2}\langle\lambda\lambda\rangle_{\mathrm{susy}} = 
 \mathrm{e}^{K/2} M^3 \mathrm{e}^{-24\pi^2 f (S,\chi)/b_0} .
\end{equation*}
Substituting this into the expression for the scalar potential one
finds that contributions coming from condensates are proportional to
$\partial_{\chi}W$ and $\partial_{S}W$ with $W$ given by
eq.(\ref{hcondensate}) --- as is indeed expected.

\subsection{The racetrack superpotential}

We are now in a position to write down the form of the superpotential
corresponding to two gaugino condensates driven by two hidden sectors
with gauge group $SU (N_1)$ and $SU (N_2)$ respectively. For
simplicity we allow for a moduli dependence in the second condensate
only. The race-track superpotential has the form
\begin{equation}
W_{\mathrm{npert}} = A N_1 M^3 \mathrm{e}^{-S/N_1} - B N_2 M^3
 \mathrm{e}^{-S/N_2} \left(\frac{M^2}{(\alpha +
 \beta \chi)^2}\right)^{3 (N_2^\prime - N_2)/(2N_2)} .  
\label{spot1}
\end{equation}
The coefficients $A,\,B$ are related to the remaining string
thresholds $\Delta_i$ by $A = \mathrm{e}^{-\Delta_1/(2N_1)}/N_1$
etc, whose moduli dependence we do not consider here.

\section{The racetrack scalar potential}

In this paper we will consider a simple model with the dilaton, $S$,
the moduli field $\chi$, and the breathing mode of the compact 6-dim
space, $T$ which is associated with the overall volume of the compact
manifold, $\mathrm{Re} T = V^{1/3}$. We adopt the following standard
K\"{a}hler potential
\begin{equation} 
K (S, \bar{S}; T, \bar{T}; \chi, \bar{\chi}) = -3 \log (T + \bar{T})
 -\log (S + \bar{S}) + \chi \bar{\chi}\ ,
\label{kah1}
\end{equation}
where $S = s + i\phi$, $\chi = x\,\mathrm{e}^{i\theta}$, $T = t +
i\eta$. An important issue is the freezing of the volume modulus
$T$. In particular for the racetrack potential alone there is a
runaway direction with $t$ tending to zero. This means $t$ must be
fixed by another sector of the theory so we address this issue first.

\subsection{Fixing the volume modulus $T$}

For the reason just discussed we concentrate here on the $T$ modulus
although much of the follwing discussion applies also to other string
moduli. The general comment is that, although moduli do not appear on
their own in the superpotential and are associated with flat
directions, these are lifted after supersymmetry breaking so the
moduli will be fixed. In this sense, provided one has a mechanism for
supersymmetry breaking, it is not necessary to fix moduli by
fluxes. As we now discuss it is not even necessary for supersymmetry
breaking to occur for moduli to acquire masses in the absence of
fluxes. This can readily happen if there is a stage of spontaneous
symmetry breaking in which non-moduli fields acquire vevs which can
destroy the flatness of the moduli potential. This is explicitly
illustrated below for the $T$ modulus.

In the large volume limit the K\"{a}hler potential for the $T$ field
assumes the form $K(T, \bar{T}) = -3 \log (T + \bar{T})$, and that is
the form which we have assumed here. The result is that the scalar
potential, except the $D$-term contribution, is multiplied by the
factor $(2t)^{-3}$. In fact, we do not need to address in detail the
issue of low-energy stabilization of $T$ --- all we require is that
during inflation $t$ should be frozen. There are a number of
mechanisms which can do this. The obvious and easiest thing in the
present context is if a matter field, $\Phi_1$, charged under an
anomalous $U(1)$ gauge group has a $T$-dependent kinetic term. This is
the case for example for fields carrying non-zero modular weights in
heterotic compactifications. Then there appears a $T$-dependent
$D$-term contribution to the scalar potential
\begin{equation}
V_D = \frac{g^2}{2} (f_1 (t)|\Phi_1|^2 - \xi)^2 ,
\end{equation}
where $\xi$ arises from the Green-Schwarz term. To preserve
supersymmetry, $\Phi_1$ must acquire a vev, and this in turn generates
a mass for $t$.  However it is clear that this term only requires
$\langle f_{1} (t)|\Phi_1|^2\rangle = \xi$, so additional terms must
be included to fix the relative value of $\langle\,t\rangle$ and
$\langle\Phi_1\rangle.$ This can readily happen through the soft mass
term $m_{\Phi_1}^2\left\vert\Phi_1\right\vert^2$. The resulting
potential has the form
\begin{equation*}
V (T, S, \chi) = V_{\mathrm{R T}} (S, \chi) \frac{(1 +
 \left\vert\Phi_1\right\vert^2)}{t^3} + \frac{g^2}{2}(f_1 (t)| \Phi_1|^2 -
 \xi)^2 ,
\end{equation*}
where $V_{\mathrm{RT}}(S, \chi)/t^3$ is the full racetrack potential
and the term $V_{\mathrm{RT}}(S,
\chi)\left\vert\Phi_1\right\vert^2/t^3$ is the
$m_{\Phi_1}^2\left\vert\Phi_1\right\vert^2$ term during inflation
which is automatically present in the supergravity potential. Clearly
minimising this with respect to $t$ and $\Phi_1$ fixes them
independently of the minimisation with respect to $S$ and $\chi$. Note
that this means we do not encounter the difficulties found in
ref.\cite{Choi:2004sx} in their analysis of flux stabilised models
where the minimisation must be done simultaneously. For the case
$\Phi_1$ has a nontrivial modular weight $n_1$ we have $f_1 =
(2t)^{-n_1}$ and hence $\langle\,t\rangle \simeq \left(\frac{3}{2\xi
(n_1 - 3)}\right)^{1/n_1}$ for $n_1 > 3$. (For $n_1 < 3$ there is
still a minimum at finite $t$ provided the soft mass $m_{\Phi_1}^2$ is
driven negative by radiative corrections at some scale.)

This example illustrates how the $T$ field can be fixed by kinetic
terms.  Another well known example with $T$-dependent kinetic terms of
matter fields are models with a warp factor. In the case of an
exponential warp factor the K\"{a}hler potential takes the form $K =
-3\log (1 - \exp(-T - \bar{T} + |\Phi|^2) - |C|^2)$ where $\Phi$
denotes warped and $C$ unwarped matter fields. If under a gauge
symmetry $T$ transforms nonlinearly, $\delta T = i q\epsilon$, and
$\Phi$ and $C$ linearly, $\delta\Phi = i\epsilon Q_{\Phi}\Phi,\,
\delta C = i\epsilon Q_{C}C$, then vanishing of the corresponding
D-term leads to the approximate relation \cite{Ciesielski:2002fs}:
\begin{equation}
\mathrm{e}^{-2t} = \frac{Q_C |C|^2}{q - Q_\Phi |\Phi|^2}\ .
\end{equation}

So far we have considered the case when there is no $T$ dependence in
the superpotential. This has the advantage that the minimum of the
potential is at vanishing cosmological constant. When the
superpotential depends on $T$, which is often the case, the scalar
potential becomes more complicated, and the actual minimum corresponds
to negative cosmological constant. This is the situation when one uses
target space modular symmetry covariant superpotentials, or when one
invokes an exponential superpotential of the form
$W_{\mathrm{nonpert}} = A N_1 \mathrm{e}^{-S/N_1} + B N_2
\mathrm{e}^{-T/N_2} - W_0$. However, even in such cases it if often
possible to argue that the largest term in the potential in the
inflationary regime is given by the one we consider, $\mathrm{e}^{K}
K^{S\bar{S}}|D_{S}W|^{2}$, with $T$ frozen, for instance by the
$D$-term as discussed above.

In fact, the rank of the gauge group permitting, one can stabilise
both $S$ and $T$ with the help of condensates and use the inflationary
mechanism discussed below to trap both of these moduli. Moreover, if
there is a second anomalous $U(1)$ with an assignment of charges such
that the Green-Schwarz term $\xi$ cannot be compensated, one has a
source of a residual cosmological constant. This may be
phenomenologically relevant provided that the moduli dependence leads
to a dramatic suppression of its magnitude.  Moduli dependent
Green-Schwarz terms are associated at the level of the effective 4-dim
Lagrangian with gaugings of the imaginary shifts of various moduli
fields, $M$. In this case the moduli dependence arises because the
Green-Schwarz terms are proportional to $\frac{\partial\,K (M,
\bar{M})}{\partial\bar{M}}$ \cite{Ciesielski:2002fs}.~\footnote{For
instance, if $K=-3\log (T + \bar{T})$ then $\xi \sim \frac{3}{2t}$, so
if $K=(\bar{M} + M)^{2}$ then $\xi \sim (\bar{M} + M)$, and the D-term
contributions to the potential are $\sim 1/t^2$ and $\sim (\bar{M} +
M)^2$ respectively.}

\subsection{The racetrack potential}

In what follows we assume that one of the mechanisms discussed in the
last Section fixes $t$ and allows us to consider the dependence of the
racetrack potential only on $S$ and $\chi$. In addition we introduce
the notation $\epsilon = (N_1 -N_2)/(N_1 N_2)$ (assuming $N_1 > N_2$)
and $\gamma = 3(N_2^\prime - N_2)/(2N_2)$.

For the case of the $T$ independent racetrack superpotential we obtain a
positive semi-definite scalar potential for the dilaton and $\chi$-modulus
given by 
\begin{eqnarray}  
& V (S) = \frac{1}{2s} \kappa \left\vert A (2 s +
 N_1)\mathrm{e}^{-s/N_1} - B \mathrm{e}^{-i \epsilon \phi} \left(\alpha +
 \beta \chi\right)^{-2 \gamma} (2 s + N_2)\mathrm{e}^{-s/N_2}\right\vert^2
 \mathrm{e}^{|\chi|^2}& \label{twocondsfin} \\ 
& + \frac{|\chi|^2}{2s} \kappa \left\vert A N_1 \mathrm{e}^{-s/N_1} - B
 \mathrm{e}^{-i \epsilon \phi}\left(\alpha + \beta \chi\right)^{-2 \gamma}
 N_2 \mathrm{e}^{-s/N_2}\left(1 - \frac{2 \gamma
 \beta}{\alpha \bar{\chi} + \beta |\chi|^2}\right) \right\vert^2
 e^{|\chi|^2},& \notag
\label{model1}
\end{eqnarray}
where $\kappa = 1/(8t^3)$ and the factor $e^{|\chi|^2}$ (in Planck
units) comes from the factor
$\mathrm{e}^{K}$ present in the supergravity potential.

After expanding the perfect squares one obtains a rather complicated
function of 4 real variables:
\begin{eqnarray}
V (s, \phi, x, \theta) &=& \frac{e^{ x^2}}{2s} \kappa 
 \left(A^2 (2s + N_1)^2 \mathrm{e}^{-2s/N_1} 
 + B^2 (2s + N_2)^2 \mathrm{e}^{-2s/N_2} 
 [r (x, \theta)]^{-4\gamma} \right. \notag \\
 && -2AB (2s + N_1)(2s + N_2) \mathrm{e}^{-s(\frac{1}{N_1} + 
 \frac{1}{N_2})} [r (x, \theta)]^{-2\gamma} 
 \cos[\epsilon \phi + 2\gamma \delta (x, \theta)]
 \left. \right) \notag \\
 &&+ \frac{e^{x^2}}{2s}\kappa \left(x^2 A^2 N_1^2 \mathrm{e}^{-2s/N_1} +
 B^2 N_2^2 \mathrm{e}^{-2s/N_2} [r (x, \theta)]^{-4\gamma} [r^\prime 
 (x, \theta)]^2 \right. \notag \\
 &&- 2 x A B N_1 N_2 \mathrm{e}^{-s(\frac{1}{N_1} + \frac{1}{N_2})}
 [r (x, \theta)]^{-2\gamma} r^{\prime} (x, \theta) \cos [\epsilon \phi +
 2 \gamma \delta (x, \theta) - \delta^\prime (x, \theta)] \left . \right),
\label{full}
\end{eqnarray}
where 
\begin{eqnarray}
& r^2 (x, \theta) = [\alpha + \beta x \cos\theta]^2 + \beta^2 x^2
 \sin^2\theta , & \notag \\
& r^\prime{}^{2} (x, \theta) = \left ( x - 2 \gamma \beta \frac{\beta x +
 \alpha \cos\theta}{\beta^2 x^2 + \alpha^2 + 2 \alpha \beta x \cos\theta}
 \right)^2 + \frac{4 \gamma^2 \beta^2 \alpha^2 \sin^2\theta}{(\beta^2
 x^2 + \alpha^2 + 2 \alpha \beta x \cos\theta)^2} \ , &  \notag \\
& \tan [\delta (x, \theta)] = \frac{\beta x \sin\theta}{\alpha + \beta
 x\cos\theta}\ , &  \notag \\
& \tan [\delta^\prime (x, \theta)] = \frac{2 \gamma \beta \alpha
 \sin\theta}{\beta^2 x^2 + \alpha^2 + 2 \alpha \beta x \cos\theta} 
 \left ( x - 2 \gamma \beta \frac{\beta x + \alpha \cos\theta}{\beta^2
 x^2 + \alpha^2 + 2 \alpha \beta x \cos\theta} \right)^{-1} \ .&
\end{eqnarray}
As discussed above it is consistent to fix $T$ and $\kappa$
independently of the minimisation with respect to $S$ and $\chi$. Thus
in the above potential $T$ and $\kappa$ are to be considered
constant. Their values provide additional parameters which determine
the overall scale of the potential.

\subsubsection{Properties of the pure dilaton potential \label{pure}}

Given the complexity of the potential it is very difficult to study
the most general form as a function of the 4 real variables. As a
start to understanding its structure we first study the properties of
the purely dilatonic potential.~\footnote{We put $\kappa = 1$ and $M =
1$ unless stated otherwise.} While we will find that it can generate
inflation it suffers from the problem of dilaton overshoot after
inflation leading to the non-interacting theory in the
post-inflationary era. Thus this Section will serve to motivate the
consideration of the more involved potential involving the moduli. The
purely dilatonic potential has the form:
\begin{equation}
V (S) = \frac{1}{2s} \left\vert A(2s + N_1)\mathrm{e}^{-s/N_1} \mathrm{e}
^{i\epsilon\phi} - B(2s+N_2) \mathrm{e}^{-s/N_2}\right\vert^2.
\label{twoconds}
\end{equation}
This can be written down as a sum of two positive semi-definite terms 
\begin{eqnarray}
& V (s, \phi) = \frac{1}{2s}\left(A(2s + N_1)\mathrm{e}^{-s/N_1} - B(2s +
 N_2) \mathrm{e}^{-s/N_2}\right)^2 & \notag  \label{poten1} \\
& + \frac{1}{s} A B (2s + N_1)(2s + N_2)\mathrm{e}^{-(s/N_1 + s/N_2)} 
 \left(1 - \cos(\phi \epsilon)\right) .&
\end{eqnarray}
Only the second term depends on the phase $\phi$, and it is minimised
for $\phi_k = k\,\pi/\epsilon,\;k = 0, \pm 1, \pm 2, \ldots$ Notice
that the minimum corresponds to a relative minus sign between the two
condensates which allows for the cancellation between the terms. This
can lead to one or more minima at finite $s$, but, as mentioned above,
it may be very difficult to access these minima. For example if
initially the phase is displaced from its minimum the cancellation
between the terms is no longer complete and the dilaton may be in
domain of attraction of the run-away region $s\rightarrow\infty$. We
will return in later Sections to the question of how to access minima
of the racetrack potential at finite $s$.

It is straightforward to establish that, for $B/A>1$, there is a
minimum of the potential at finite $s$ and that the potential vanishes
there. We will discuss below whether inflation occurs in the flow from
some initial value of $s$ towards this minimum. If $1>B/A>N_2/N_1$,
there is still a minimum at finite $s$ but the potential does not
vanish there. Although for the pure dilaton potential this de Sitter
minimum is not suitable for generating a finite period of inflation,
its existence may be relevant to the case the superpotential has
additional moduli dependence allowing for an exit from the de Sitter
phase. We will explore this possibility below. Note that in general
the first term in the dilaton potential may have up to 4 finite
critical points, and one more with infinite vacuum value of the
dilaton.

\section{Challenges for racetrack inflation}

As mentioned in the Introduction, racetrack models have been
extensively explored earlier and several difficulties identified which
have so far prevented the construction of a viable model (in the
absence of fluxes) which can yield inflation and lead to an acceptable
Universe afterwards. In this Section we briefly review these
difficulties.

\subsection{The weak coupling regime of the pure dilaton potential}

We first discuss the possibility that, in the pure dilaton potential
of eq.(\ref{twoconds}) there is inflation in the flow to the minimum
at finite $s$ in the weak coupling limit. In this limit and assuming
the phase of the dilaton field is at its minimum, the position of the
minimum and the position of the maximum of the barrier separating it
from the minimum at infinity are given by
\begin{equation}
s_{\mathrm{min}} = \frac{1}{\epsilon }\log\left(\frac{B}{A}\right) , \quad
s_{\mathrm{max}} = \frac{1}{\epsilon}\log\left(\frac{BN_1}{AN_2}\right) .
\end{equation}
It is straightforward to check whether the necessary conditions for
slow-roll inflation in the $s$ direction are fulfilled at the top of
the barrier, i.e. whether $\eta = s^2\frac{\partial^2 V}{\partial s^2}
\ll 1$. In the large $s$ limit this can be calculated to be
\begin{equation}
\eta = 2s_{\mathrm{max}}^{2}\frac{\log(\frac{BN_1}{AN_2})}{N_1 - N_2} ,
\end{equation}
which is larger than unity in the weak coupling domain, thus slow-roll
inflation does not occur.

The other possibility for the pure dilaton potential in the weak
coupling domain is slow-roll inflation in the direction of the phase
of the dilaton.  The slow-roll parameter is now given by: $\eta_\phi =
(\epsilon s)^2 \frac{\cos(\epsilon\phi)}{1 -
\cos(\epsilon\phi)}$.~\footnote{Here we assume that the dilaton sits
at the minimum of the first term of eq.(\ref{poten1}).} The
requirement that $\eta$ should be much smaller than unity imposes a
lower bound on the value of the phase $\phi$. Unfortunately for such
high values of the phase, the minimum along the direction of the
dilaton has already disappeared --- the necessary condition for the
presence of that minimum is $\cos^2(\epsilon\phi) > 4(N_2/N_1)/(1 +
N_2/N_1)^2$, which prefers smaller values of the phase.

\subsection{The strong coupling regime and the `rapid roll problem'}

In order to evade this conclusion and to find a region where inflation
can occur it is necessary to move towards the strong coupling
domain. In fact there is a saddle-point of the potential (\ref{spot1})
in the strong coupling regime which can generate a significant amount
of inflation.~\footnote{Of course in the strong coupling domain there
will be significant corrections to the potential given by
eq.(\ref{twoconds}) so finding a region of $s$ and $\phi$ where
inflation does occur using the potential (\ref{spot1}) can only be
considered as indicative.} However now we face the fundamental problem
of explaining how, after inflation, the dilaton moves to the weak
coupling minimum at finite $s$ rather than to the non-interacting
region $s\rightarrow \infty$. Since the problem occurs at weak
coupling it can be reliably estimated even though inflation occurs at
strong coupling.

To illustrate the magnitude of the problem we consider a particular
example of eq.(\ref{twoconds}) with $\gamma = 3(N_2^\prime -
N_2)/(2N_2) = 1/2$, $N_1 = 10,\,N_2 = 9,\,A = 1$, $B=1.112$. The
potential (see Fig.~1) has two minima in the dilaton direction in the
strong coupling regime and there is a significant amount of inflation
generated in the roll to the minima from the local directional maximum
in the direction of $s (z)$.

\begin{figure}[htb]
\begin{center}
\epsfig{file=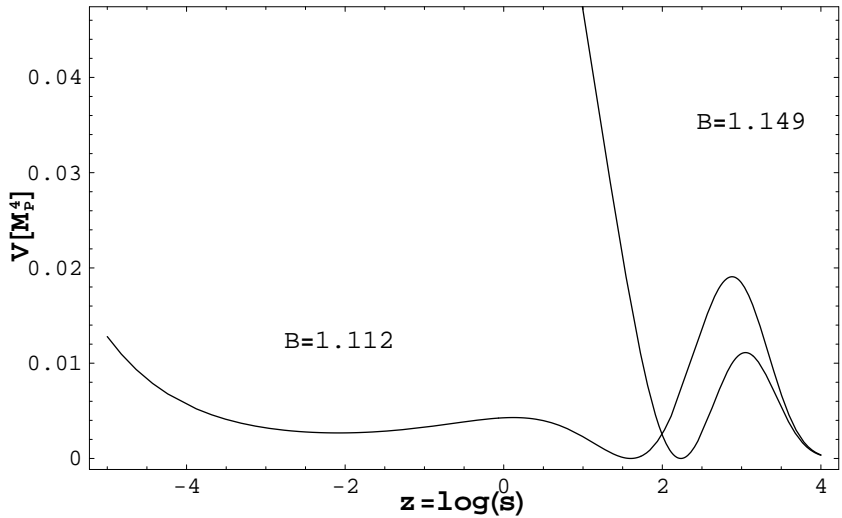, width=.65\linewidth} 
\label{f1} 
\end{center}
\par
\caption{The shape of the potential in the dilaton $z$ direction (taking
$\protect\chi = \protect\alpha = 0$) for two illustrative cases: 
(i) $B = 1.112$ --- two minima, (ii) $B = 1.149$ --- a single minimum 
with vanishing energy (strongly coupled regime).}
\end{figure}

Of course it is necessary to move from the strong coupling regime at
the end of inflation and this can be done by allowing for a moduli
dependence of the coefficient $B\rightarrow B\left( 1+\beta \chi
\right) ^{-2\gamma }.$ We should now use the full potential
(\ref{twocondsfin}) but this still has the relevant saddle-point
needed to generate inflation. We now consider the dilaton dependence
of the two-condensate potential when the additional field $\chi$
varies. It does so to reduce the vacuum energy and this drives the
dilaton to larger values (see Fig.~2), where the form of the potential
(\ref{twocondsfin}) is reliable.

\begin{figure}[htb]
\begin{center}
\epsfig{file=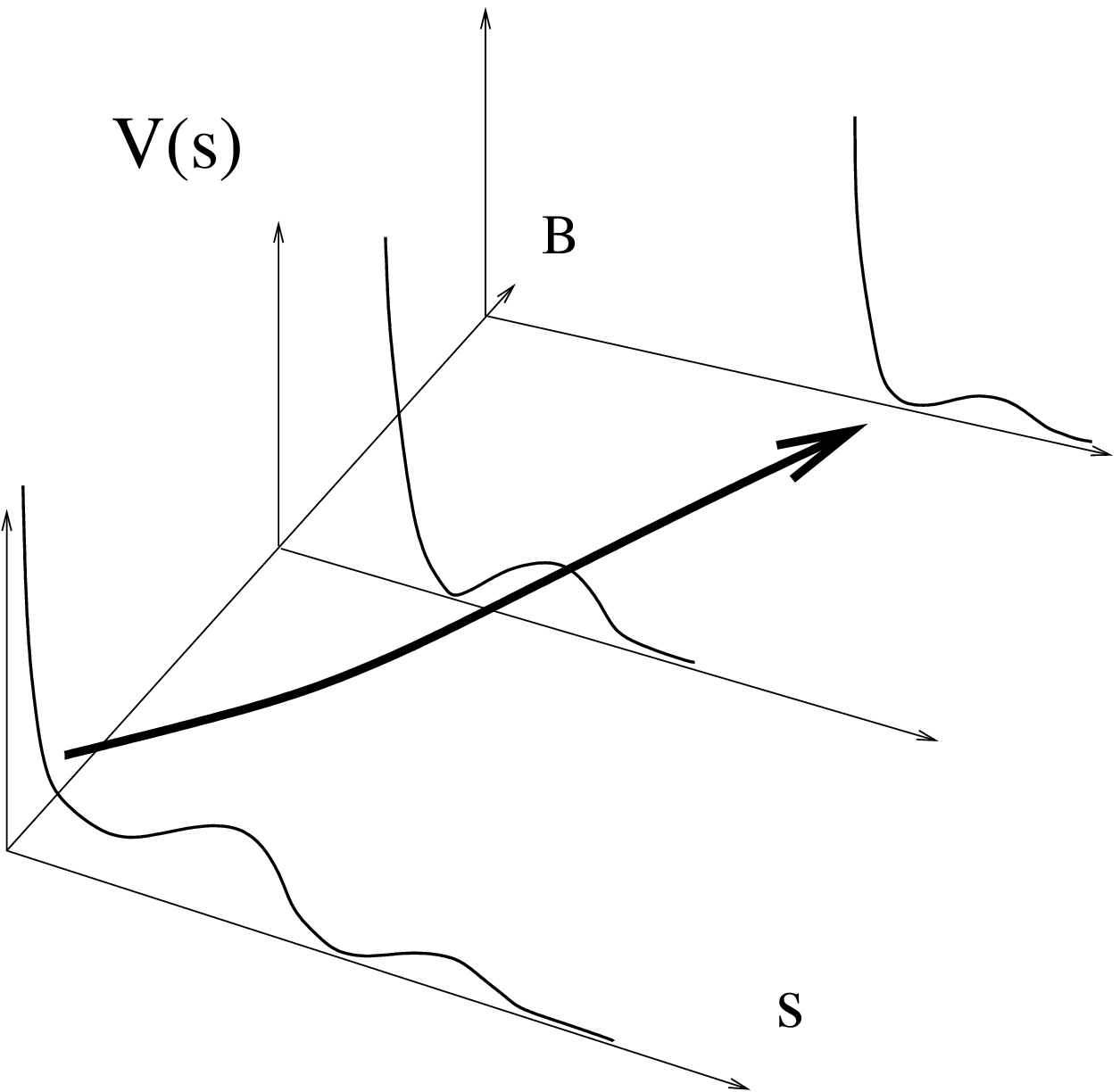, width=.5 \linewidth} 
\label{f2} 
\end{center}
\par
\caption{Sketch of how small change of parameters (represented here by
$B$) of the racetrack potential interpolates between the strongly
coupled (small $s$) and weakly coupled (large $s$) regimes
(arbitrary scales).}
\end{figure}

In what follows we switch from $s$ to the canonically normalised
variable $z = \log(s)$. The problem is that the height of the barrier
between the weakly coupled minimum and infinity is very small. We have
approximately
\begin{equation}
V_{\mathrm{weak\,barrier}} = m_{3/2}^2 \left[\frac{8\pi}{g^2} +
\frac{1}{\epsilon}
\log\left(\frac{N_1}{N_2}\right)\right]\frac{8\pi}{g^2} ,
\end{equation}
which for the parameters assumed here is of the order of $10^{-26}$ in
Planck units. At the same time it can be seen from Fig.~1 that the
vacuum energy during inflation is of order $10^{-2}$ in the same
units. Thus the vacuum energy being released is $10^{24}$ times
greater than the potential barrier. Even though the fields are damped
by expansion during the roll to weak coupling much of this potential
energy is converted to kinetic energy and it is overwhelmingly likely
that the dilaton will simply jump the barrier between the weakly
coupled minimum and infinity and flow to the unphysical decoupling
limit. This is the rapid-roll problem emphasised by Brustein and
Steinhardt \cite{Brustein:1992nk}.

The underlying problem is that the scale of racetrack potential is
dominated by the exponential factors $\mathrm{e}^{-s/N_{1,2}}$, and in
the roll from strong to weak coupling these factors require that there
is a huge release of potential energy. To avoid this problem it is
necessary to look for an inflationary regime which occurs at weak
coupling. This offers the possibility of evading the rapid-roll
problem because the barrier between the physical minimum and the
non-interacting minimum is of order of the vacuum energy during
inflation. As discussed above weak coupling inflation does not happen
for the pure dilaton potential but we will show that it can occur in
the moduli dependent racetrack potential.

\subsection{Finite temperature effects and the `thermal roll problem'}

A related problem to the rapid roll problem is the possibility that
thermal effects will drive the dilaton to the non-interacting region
$s\rightarrow \infty .$ This has been discussed recently by
Buchm\"{u}ller \emph{et al.} \cite{Buchmuller:2004xr}. The point is
that since the couplings of hot matter depend on the dilaton, the free
energy of the hot gas adds to the zero-temperature effective potential
for the dilaton, and this may change its behaviour if the
temperature-dependent piece is sufficiently large. More specifically,
the complete dilaton potential is
\begin{equation}
V_{\mathrm{tot}} = V (s) + F (g, T) = V (s) - \frac{\pi^2 T^4}{24} 
 \left(a_0 + a_2 g^2 + \mathcal{O}(g^3)\right),
\end{equation}
where the hot QCD free energy has been taken into account and the
gauge coupling $g$ is a function of the dilaton $s$. For the hot QCD
plasma, the coefficient $a_0$ is positive, but $a_2$ is negative so
the second term has a minimum for $g = 0$. For temperatures above a
critical value, $T_{\mathrm{crit}}$, this term fills in the minimum of
the $T = 0$ dilaton potential at finite $s$ and as a result at high
temperatures the dilaton is driven to the non-interacting regime and
remains there as the temperature drops. The critical temperature was
estimated to be of $\mathcal{O}(10^{13}$) GeV
\cite{Buchmuller:2004xr}. In the context of inflation driven by the
racetrack potential, the problem is that such thermal effects can move
the initial conditions far from the saddle point at which inflation
occurs.

However there is a straightforward possibility that evades this
problem. At a very high value for the Hubble parameter, close to the
Planck scale, there is insufficient time for scattering processes to
establish thermal equilibrium \cite{Ellis:1979nq}. In this era the
dilaton will not feel the thermal potential and can be in the region
where inflation occurs. To quantify this we note that the interaction
rate in the quark-gluon gas can be approximated as
\begin{equation}
\Gamma_{\mathrm{int}} = \alpha_{\mathrm{s}}^2\,T
\end{equation}
and the expansion rate is given by $H_{\mathrm{hot}} =
\sqrt{N_{\mathrm{eff}}}T^2/M_{\mathrm{P}}$, where $N_{\mathrm{eff}}$
is the effective number of massless degrees of freedom and
$M_{\mathrm{P}} \simeq 2.4 \times 10^{18}$ GeV is the reduced Planck
scale. Requiring that $\Gamma_{\mathrm{int}} > H_{\mathrm{hot}}$
gives the condition for thermal equilibrium:
\begin{equation}
T < T_{\mathrm{eq}} = M_{\mathrm{P}}
\frac{\alpha_{\mathrm{s}}^2}{\sqrt{N_{\mathrm{eff}}}}.
\end{equation}
This means the universe cannot thermalise above $T_{\mathrm{eq}} \sim
10^{15}$\,GeV. Although a rough estimate, this is close to the value
of $T_{\mathrm{eq}} = 3 \times 10^{14}$\,GeV from a careful analysis
\cite{Enqvist:1993fm}. So provided the inflationary potential scale is
{\em above} $\sqrt{\alpha_{\mathrm{s}}^2 T_{\mathrm{eq}}
M_{\mathrm{P}}} \sim 3 \times 10^{15}$~GeV, the thermal-roll problem
is avoided.~\footnote{Note that this scale must also be {\em below}
$\sim 2\times10^{16}$~GeV in order to respect the observational upper bound
from WMAP-3 on gravitational waves generated during inflation
\cite{Kinney:2006qm}.}  Of course a viable
theory must ensure that the thermal-roll problem does not reappear
after inflation in the reheat phase. This simply puts an upper bound
on the reheat temperature, $T_{\mathrm{rh}},$ given by
$T_{\mathrm{rh}} < T_{\mathrm{eq}}$. This is however far less
stringent than the upper bound of $T_{\mathrm{rh}} <
10^{8}-10^{10}$\,GeV set already by consideration of gravitino
production \cite{Ellis:1984er}, so this will not be a problem in any
phenomenologically acceptable model.

\section{Inflation from the racetrack potential}

We will now argue that the racetrack potential has all the
ingredients to meet the challenges just discussed. There are two main
aspects to this. Firstly the domain walls generated by the racetrack
potential naturally satisfy the conditions needed for topological
inflation.  As a result there will be eternal inflation within the
wall which sets the required initial conditions for a subsequent
period of slow-roll inflation during which the observed density
perturbations are generated. The second aspect is the existence of a
saddle point(s) close to which the potential is sufficiently flat to
allow for slow-roll inflation in the weak coupling domain. As noted
above this is not the case for the pure dilaton potential but does
occur when one includes a simple moduli dependence.

\subsection{Topological inflation}

As pointed out by Vilenkin \cite{Vilenkin:1994pv} and Linde
\cite{Linde:1994wt}, ``topological inflation'' can occur within a
domain wall separating two distinct vacuua. The condition for this to
happen is that the thickness of the wall should be larger than the
local horizon at the location of the top of the domain wall (we call
this the `coherence condition'). In this case the initial conditions
for slow-roll inflation are arranged by the dynamics of the domain
wall which align the field configuration within the wall to minimise
the overall energy. The formation of the domain wall is inevitable if
one assumes chaotic initial conditions which populate both distinct
vacuua and moreover walls extending over a horizon volume are
topologically stable. Although the core of the domain wall is stable
due to the wall dynamics and is eternally inflating, the region around
it is not. As a result there are continually produced regions of space
in which the field value is initially close to that at the centre of
the wall but which evolve to one or other of the two minima of the
potential. If the shape of the potential near the wall is almost flat
these regions will generate a further period of slow-roll inflation,
at which time density perturbations will be produced.

In the case of the racetrack potential the coherence condition
\cite{Vilenkin:1994pv} necessary for topological inflation appears to
be rather easily satisfied. This condition states that the physical
width of the approximate domain wall interpolating between the minimum
at infinity and the minimum corresponding to a finite coupling should
be larger than the local horizon computed at the location of the top
of the barrier that separates them. The width of the domain wall,
$\Delta$, is such that the gradient energy stored in the wall equals
its potential energy, $\left(\frac{2\delta}{\Delta}\right)^2 = V
(s_{\mathrm{max}})$, where $\delta = \log
(s_{\mathrm{max}}/s_{\mathrm{min}})$.~\footnote{One should use here
the canonically normalised variable $z = \log (s)$.} For the racetrack
potential (\ref{twoconds}), the ratio of the width to the local
horizon --- the coherence ratio --- turns out to be:
\begin{equation}
\Delta/H^{-1} = \sqrt{32\pi/3} \log\left(1 + N_1 N_2
\log(N_1/N_2)/s_{\mathrm{min}}\right) .
\end{equation}
The weak coupling region corresponds to $s > 2\pi$.~\footnote{We are
neglecting the string thresholds.} Since $s_{\mathrm{min}} \simeq N_1
N_2$ one sees that the coherence ratio is typically larger than 1 in
this region.

In fact one can obtain another form for the coherence ratio that
demonstrates this more clearly. Close to the top of the domain wall
one can approximate the potential in its neighbourhood by the zeroth
and quadratic terms in the Taylor expansion: $V (z) = V_0 -
\frac{\mu^2}{2}z^2$. Let us define the value of $z$ limiting the
domain wall as $V(z_{1/n}) = V_0/n$.  Then noticing that the spatial
width of the domain wall is such that the potential energy is
comparable to the gradient energy $V_0 = (2z_{1/n})^2/\Delta^2$ one
finds that the ratio $\Delta/H^{-1}$ is inversely proportional to the
square root of the slow-roll parameter $\eta = V^{\prime\prime}/V$:
\begin{equation}
\frac{\Delta^2}{H^{-2}} = \frac{32\pi}{3}\frac{2(n-1)}{n}\frac{1}{\eta} .
\end{equation}
Thus even for $\eta$ of $\mathcal{O}(1)$ the condition for topological
inflation is satisfied and even more comfortably so if $\eta$ is as
small as is needed for an acceptable spectral index for density
perturbations during the slow-roll inflationary period discussed
below.

The fact that the racetrack potential readily generates topological
inflation offers solutions to {\em all} the problems discussed
above. With chaotic initial conditions for the dilaton at the Planck
era the different vacuua will be populated because the height of the
domain walls separating the minima is greater than $T_{\mathrm{eq}}$
so thermal effects will not have time to drive the dilaton to large
values. This avoids the initial thermal-roll problem. Then there will
be regions of space in which the dilaton rolls from the domain wall
value into the minimum at finite $s$, moving from larger to smaller
values, and thus avoiding the thermal-roll problem. In fact once
created the vacuum bag at finite $s$ is stable, because it cannot move
back into the core and over to the other vacuum --- the border of the
inflating wall escapes exponentially fast --- so the respective region
of space is trapped in the local vacuum. Finally, if after inflation
the reheat temperature is lower than the critical temperature
$T_{\mathrm{crit}}$ the thermal effects will be too small to fill in
the racetrack minimum at finite $s$ and the region of space in this
minimum will remain there, thus avoiding a late thermal-roll problem.

Although the existence of topological inflation seems necessary for a
viable inflationary model, by itself it is not sufficient to generate
acceptable density perturbations. What is needed is a subsequent
period of slow-roll inflation with the appropriate characteristics to
generate an universe of the right size ($>50-60$ e-folds of inflation)
and density perturbations of the magnitude observed. In the next
Section we show that the racetrack potential has the correct
properties to achieve this if one includes a simple moduli dependence
of the form discussed in Section~\ref{modulidependence}.

\subsection{Inflation in the weak coupling regime}

In this Section we present an example of a viable inflationary model
in which the inflaton is the pseudo-Goldstone boson associated with
the phase $\theta $ of the field $\chi$.

\begin{figure}[tbh]
\begin{center}
\epsfig{file=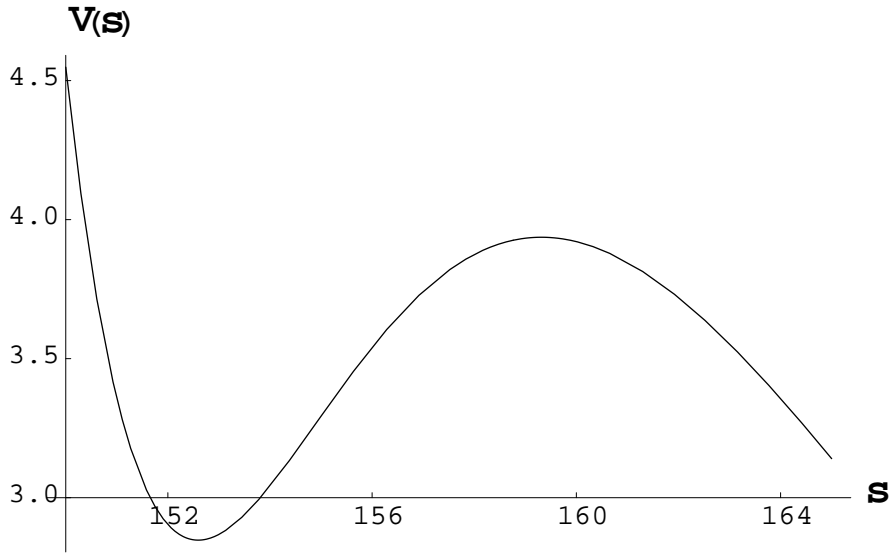, width=.7\linewidth} 
\label{f2.1} 
\end{center}
\par
\caption{The $s$ dependence of the potential (in units of $10^{-15} M_\mathrm{P}^4$) 
in the neighbourhood of the weakly coupled minimum.}
\end{figure}

\begin{figure}[htb]
\begin{center}
\epsfig{file=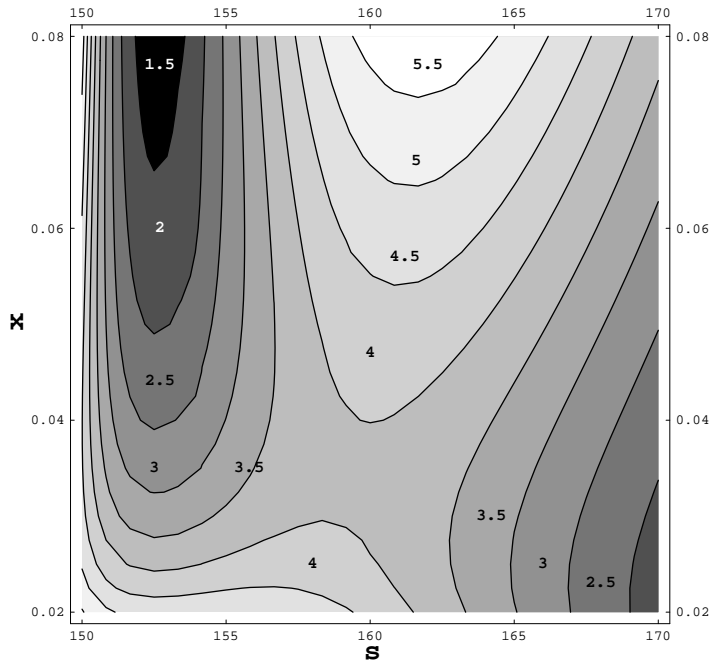, width=.55 \linewidth} 
\label{f2.2} 
\end{center}
\par
\caption{The contour plot in the $s,x$ plane in the neighbourhood of
the saddle point. The numbers on the contours give the corresponding values 
of the scalar potential in units of $10^{-15} M_\mathrm{P}^4$.}
\end{figure}

\begin{figure}[htb]
\begin{center}
\epsfig{file=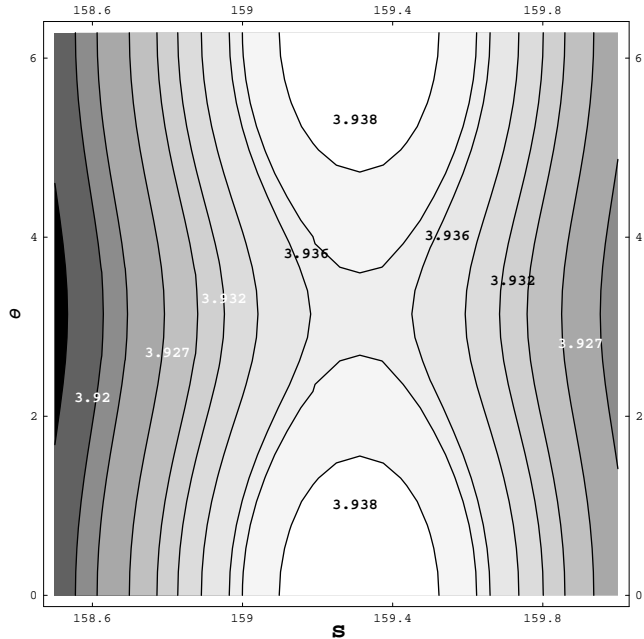, width=.5 \linewidth} 
\label{f2.3} 
\end{center}
\par
\caption{The $\protect\theta$ dependence of the potential in the
neighbourhood of the saddle point. Note that the slope {\em along} the
direction of $\theta$ is much smaller than the slope in the direction
of $s$. The numbers on the contours give the corresponding values 
of the scalar potential in units of $10^{-15} M_\mathrm{P}^4$.}
\end{figure}

\begin{figure}[htb]
\begin{center}
\epsfig{file=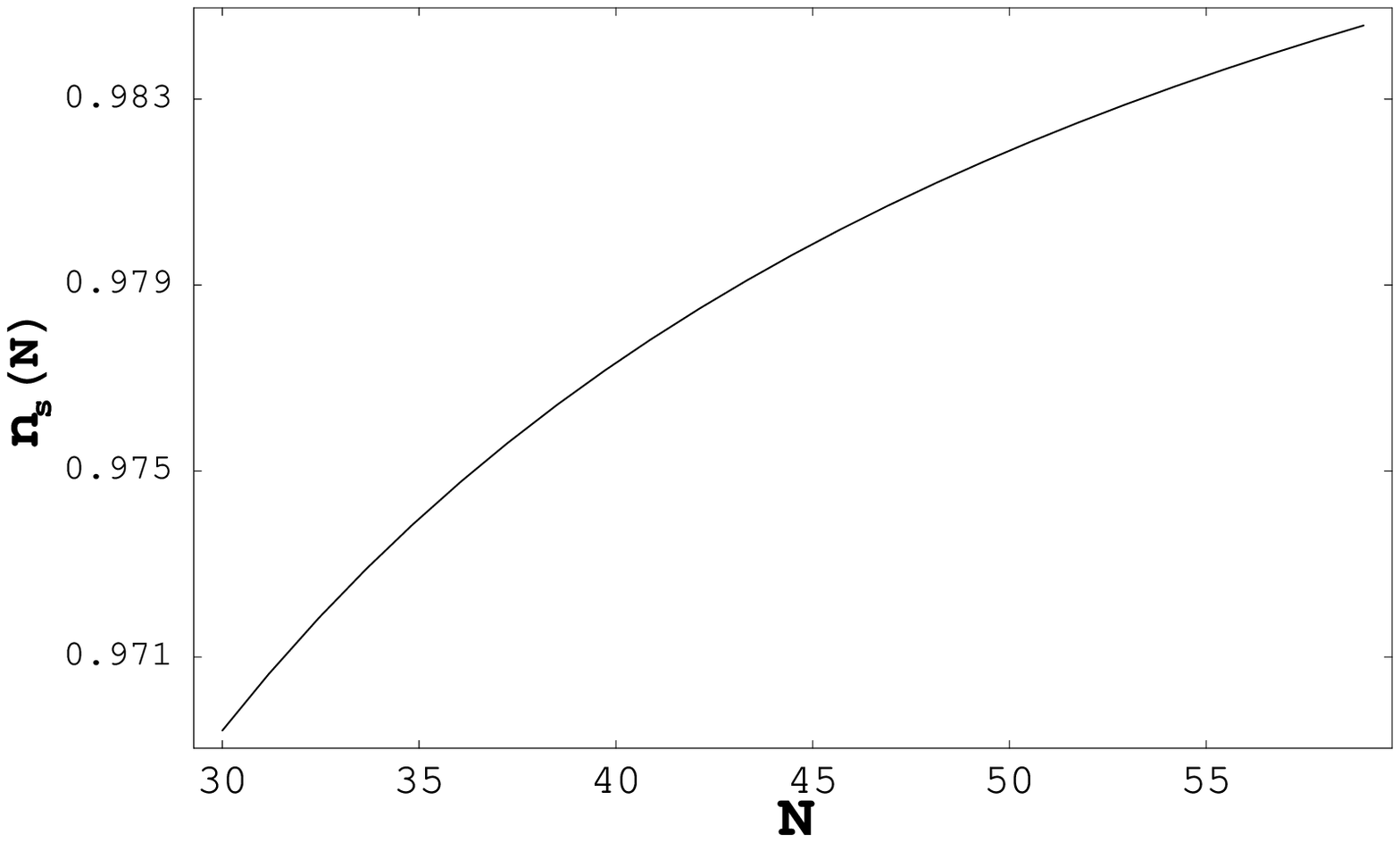, width=.7 \linewidth} 
\label{f13} 
\end{center}
\par
\caption{The ``running'' of the scalar spectral index, plotted as a
function of the number of e-folds from the end of inflation. Note that
observationally relevant scales correspond to just the last 5 or so
e-folds.}
\end{figure}

We start with the K\"{a}hler potential, eq.(\ref{kah1}). The model
relies on additional threshold factors of the type discussed in
eq.(\ref{fieldthreshold}) coming from the contribution of additional
massive states to the beta function. For simplicity we consider the
case that these states have mass given by the vev of $\chi $ up to a
coupling constant. The effect of the threshold factors is to add the
multiplicative factors $\chi ^{p},$ $\chi ^{p^{\prime }}$to the two
condensates giving the superpotential.
\begin{equation}
W_{\mathrm{npert}}=\chi ^{p}AN_{1}M^{3}\mathrm{e}^{-S/N_{1}}-\chi
^{p^{\prime }}BN_{2}M^{3}\mathrm{e}^{-S/N_{2}}\left(
\frac{M^{2}}{(\alpha +\beta \chi )^{2}}\right) ^{3(N_{2}^{\prime
}-N_{2})/(2N_{2})}.
\label{spotf}
\end{equation}
Here the exponents $p$ and $p^{\prime }$ are given by the change in
the beta functions $(b_{0}^{\prime }-b_{0})/b_{0}$, $c.f.$
$eq.(\ref{fieldthreshold})$.  In what follows we consider $p^{\prime
}=p$, although typically these powers are different for different
gauge sectors. For a range of $p$ and $p^{\prime }$ this
simplification does not make a significant difference to the
inflationary potential. The analytic expressions for the scalar
potentials are not very illuminating, but we give them below for
completeness:
\begin{eqnarray}
V (s, \phi, x, \theta) &=& \frac{\mathrm{e}^{x^2}}{2s} \kappa x^{2p}
 \left(A^2 (2s + N_1)^2 \mathrm{e}^{-2s/N_1} + B^2 (2s + N_2)^2 
 \mathrm{e}^{-2s/N_2} [r (x, \theta)]^{-4 \gamma} \right. \notag \\
&& - 2A B (2s + N_1)(2s + N_2) \mathrm{e}^{-s (\frac{1}{N_1} 
 + \frac{1}{N_2})} [r (x, \theta)]^{-2 \gamma} \cos [\epsilon \phi 
 + 2 \gamma \delta (x, \theta)]\left. {}\right) \notag \\ 
&& + \frac{e^{x^2}}{2s} \kappa x^{2p} (1 + \frac{p}{x^2})^2 
 \left(x^2 A^2 N_1^2 \mathrm{e}^{-2s/N_1} + B^2 N_2^2 \mathrm{e}^{-2s/N_2} 
 [r (x, \theta)]^{-4 \gamma} r^{\prime}{}^2 (x, \theta)\right. \notag \\
&& - 2 x A B N_1 N_2 \mathrm{e}^{-s (\frac{1}{N_1} + \frac{1}{N_2})} 
 [r (x, \theta)]^{-2 \gamma} r^\prime (x, \theta) \cos(\epsilon \phi 
 + 2\gamma \delta (x, \theta) - \delta^\prime (x, \theta))\left. \right) ,
\label{full'}
\end{eqnarray}
where 
\begin{eqnarray}
& r^2 (x, \theta) = [\alpha + \beta x \cos\theta]^2 + \beta^2 x^2 
 \sin^2\theta),&  \notag \\
& r^{\prime}{}^2 (x, \theta) = \left(x - 2\tilde{\gamma} (x) \beta 
 \frac{\beta x + \alpha \cos\theta}{\beta^2 x^2 + \alpha^2 
 + 2 \alpha \beta x \cos\theta}\right)^2 + \frac{4 \tilde{\gamma}^2
 (x) \beta^2 \alpha^2 \sin^2\theta}{(\beta^2 x^2 + \alpha^2 
 + 2\alpha \beta x \cos\theta)^2}\ ,&  \notag \\
& \tan [\delta (x, \theta)] = \frac{\beta x\sin\theta}{\alpha 
+ \beta x\cos\theta}\ ,& \notag \\
&\tan [\delta ^\prime (x, \theta)] = \frac{2\tilde{\gamma} (x) \beta \alpha
 \sin\theta}{\beta^2 x^2 + \alpha^2 + 2 \alpha \beta x \cos\theta}
 \left(x - 2 \tilde{\gamma}(x) \beta \frac{\beta x + \alpha
 \cos\theta}{\beta^2 x^2 + \alpha^2 + 2\alpha\beta x \cos\theta}\right)^{-1} .&
\end{eqnarray}
and $\tilde{\gamma} (x) = \gamma (1 + \frac{p}{x^2})^{-1}$. As
discussed above it is consistent to fix $T = t + i\eta$ independently
of the minimisation with respect to $S$ and $\chi $. Thus in the above
potential $\kappa = \kappa (t)$ is to be considered constant. Its value
provides an additional parameter which determines the overall scale of
the potential.

Numerical analysis of the complete Lagrangian in the $(S,\,\chi)$
hyperplane shows that typically it admits inflationary solutions. A
nice example corresponds to the choice of parameters $A = 1.5,\;B =
8.2,\;N_1 = 10,\;N_2 = 9,\;p = 0.5,\;\alpha = 1,\;\beta = 2.3$ and
$\gamma = 10^{-4}$. There is a weakly interacting minimum at $s =
152.6,\;\phi = 0,\;x=0.42$ and $\theta = 3.16$ (we remind the reader
that $S=s+i\phi$ and $\chi =xe^{i\theta}$). The structure of the
potential in the neighbourhood of $s=152.6$ is shown in Figure 3 from
which it may be seen that there is a maximum at $s = 162.2$. There is
a domain wall between the weakly interacting minimum and the
non-interacting minimum at $s=\infty $. As may be seen from Figures 4
and 5 it has a saddle point at $s=162.2,$ $\phi =0,$ $x=0.074,$
$\theta =3.152$. Inflation occurs (eternally) within the domain wall and
there is further slow-roll inflation outside the wall as it inflates
to a size not supported by the dynamics generating it.

The initial conditions for this slow-roll deserve comment. The $s$
field starts very close to the saddle point at $s = 162.2$. The same
is true for the fields $\phi$ and $x$ which, at the saddle point, have
masses larger than the Hubble expansion parameter at this
point. However the field $\theta$ is a pseudo-Goldstone field and
acquires a mass-squared proportional to $\gamma$. As may be seen from
Figure 5, since $\gamma$ is small, the potential is very flat in the
$\theta$ direction and the mass of $\theta$ is much smaller than the
Hubble expansion parameter. As a result the vev of $\theta$ during the
eternal domain wall inflation undergoes a random walk about the saddle
point and so its initial value can be quite far from saddle point.

With these initial conditions it is now straightforward to determine
the nature of the inflationary period after the fields emerge from the
region of the domain wall. This corresponds to the roll of the fields,
$s$, $\phi$ and $x$ from the saddle point to the weakly interacting
minimum, but allows for $\theta$ to be far from the saddle point. The
$s$, $\phi$ and $x$ fields rapidly roll to their minima. However the
gradient in the direction of the $\theta$ field is anomalously small
due to the pseudo-Goldstone nature of the field. Quantitatively, in
the neighbourhood of the weakly interacting minimum, we find a
negative eigenvalue of the squared-mass matrix corresponding to the
phase $\theta$, and its absolute value is about $10^4$ times smaller
than the positive eigenvalues. This is much smaller than the Hubble
expansion parameter at the start of the roll and so the $\theta$ field
indeed generates slow-roll inflation. The remaining degrees of freedom
can be integrated out along the inflationary trajectory. Inflation
stops after about $7800$ e-folds at $\theta_\mathrm{e} = 3.54$ and the
pivot point corresponds to $\theta_\star = 4.71$. The value of the
curvature of the potential at this point is $\eta_\star = -0.0089$ so
the scalar spectral index is $n_\star \simeq 1 + 2\eta_\star \simeq
0.98$, consistent with the WMAP-3 value at $2\sigma$. The agreement
with the normalisation of the spectrum is also readily achieved (we
remind the reader that the expectation value of $t$ can be considered
as a free parameter for the purpose of tuning the overall height of
the inflationary potential, as it is fixed in a separate sector of the
model). The evolution of the spectral index during inflation is shown
in Figure~6. Note that the `running' of the spectral index is very
small, $\mathrm{d}\ln n_{\mathrm{s}}/\mathrm{d}\ln k<\times 10^{-5}$,
hence probably undetectable.

To summarise, the moduli dependent racetrack potential has a saddle
point which lead to a phenomenologically acceptable period of
slow-roll inflation with the inflaton being a component of the
moduli. No fine-tuning of parameters is required and the initial
conditions are set naturally by the first stage of topological
inflation. After inflation there will be a period of reheat and the
nature of this depends on the non-inflaton sector of the theory which
we have not specified here. From the point of view of the racetrack
potential the main constraint on this sector is that the reheat
temperature should be less than $T_{\mathrm{crit}}$ to avoid the
thermal roll problem. However $T_{\mathrm{crit}}$ is quite high, much
higher for example than the maximum reheat temperature allowed by
considerations of gravitino production, so this constraint should be
comfortably satisfied in any acceptable reheating model.

\section{Summary and outlook}

There are notable differences between gaugino condensation and fluxes
as sources of the superpotential. First of all, the condensation scale
is sensitive to the actual mass spectrum at a given energy scale, so
the effective nonperturbative potential responds to expectation values
of various fields which control the actual mass terms. This readily
provides promising candidates for the inflaton field. Fluxes, on the
other hand, are imprinted into the geometry of the compact manifold
and the parameters of the respective superpotential are {\em fixed}
below the Planck scale. In fact, the coefficients of these effective
superpotentials are quantised in units of the Planck scale, which
implies that any lower energy scale can arise only through a
miscancellation of these Planck scale terms. By contrast gaugino
condensation explains the presence of the inflationary energy scale
through logarithmic renormalisation group evolution, and the tuning of
the parameters of the effective superpotential can be understood
naturally in terms of stringy and field-theoretical threshold
corrections.

The formalism developed here takes into account the backreaction of
other fields, which must be excited in the early universe, on the
dilaton. The resulting inflationary scheme has several attractive
features :

\begin{itemize}

\item There is an initial period of topological inflation which
naturally sets the initial conditions for slow-roll inflation and
avoids the rapid-roll and thermal-roll problems usually associated
with the racetrack potential. As a result there is no difficulty in
having our universe settle in the weakly coupled minimum of the
dilaton potential instead of evolving to the runaway non-interacting
minimum. This reinstates gaugino condensation as an attractive
candidate for supersymmetry breaking.

\item The racetrack potential with simple moduli dependence has saddle
points which lead to slow-roll inflation capable of generating the
observed density fluctuations with a spectral index close to
unity. Due to the initial period of topological inflation, the initial
conditions for the slow-roll inflation are set automatically without
fine-tuning. The fact that one is initially close to a saddle point
also means that there is no need to fine-tune the curvature of the
potential as is usually necessary for supergravity inflation models.

\item The inflationary models constructed here lie within the well
known field theoretical framework of gaugino-condensation induced
racetrack models. Thus the mechanism should be applicable to a wide
range of models obtained from various string-theoretical setups, {\em
including} models with fluxes. In fact, including the possible effects
of fluxes, e.g. a constant piece in the superpotential, or those of
warping of the internal space such as an exponential superpotential
for the volume modulus $T$, is rather straightforward.

\end{itemize}

\bigskip

\centerline{\Large \bf Acknowledgements}

\vspace*{0.5cm} We would like to thank Andre Lukas for useful
discussions. Z.L. thanks Theoretical Physics of Oxford University for
hospitality and S.S. acknowledges a PPARC Senior Research Fellowship
(PPA/C506205/1). This work was partially supported by the EC research and 
training networks MRTN-CT-2004-503369 and HPRN-CT-2000-00152, as well as by 
the Polish State Committee for Scientific Research (grant KBN 1 P03D 014 26) 
and POLONIUM 2005.

\end{document}